# Bornil: An open-source sign language data crowdsourcing platform for AI enabled dialect-agnostic communication

*Shahriar Elahi Dhruvo[1,2], *Mohammad Akhlaqur Rahman[1,2], Manash Kumar Mandal[2]
Md. Istiak Hossain Shihab[1,2,8], A. A. Noman Ansary[2,3], Kaneez Fatema Shithi[4], Sanjida Khanom[4]
Rabeya Akter[4], Safaeid Hossain Arib[4], M.N. Ansary[2,5], Sazia Mehnaz[2], Rezwana Sultana[2], Sejuti Rahman[4]
Sayma Sultana Chowdhury[1], Sabbir Ahmed Chowdhury[4,7], Farig Sadeque[2,3], Asif Sushmit[2,6]
farig.sadeque@bracu.ac.bd, sushmit@ieee.org
[1]Shahjalal University of Science and Technology   [2]Bengali.AI   [3]BRAC University   [4]University of Dhaka
[5]Apsis Solutions Ltd.   [6]Rensselaer Polytechnic Institute   [7]University of the West of Scotland   [8]Oregon State University

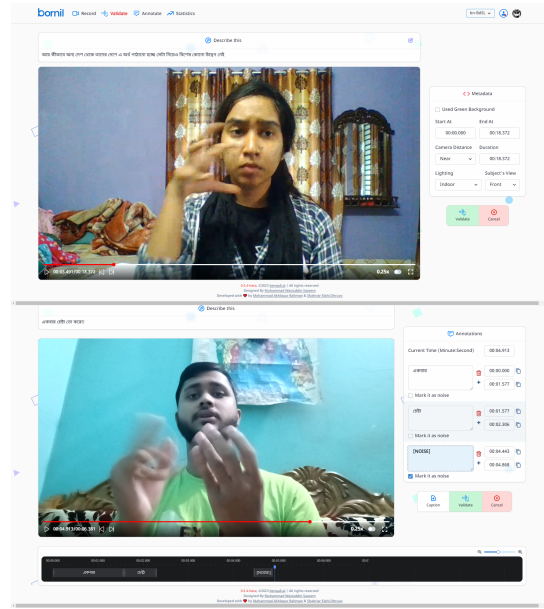

Figure 1: Snapshot of the platform

## Abstract

The absence of annotated sign language datasets has hindered the development of sign language recognition and translation technologies. In this paper, we introduce Bornil; a crowdsource-friendly, multilingual sign language data collection, annotation, and validation platform. Bornil allows users to record sign language gestures and lets annotators perform sentence and gloss-level annotation. It also allows validators to make sure of the quality of both the recorded videos and the annotations through manual validation to develop high-quality datasets for deep learning-based Automatic Sign Language Recognition. To demonstrate the system's efficacy; we collected the largest sign language dataset for Bangladeshi Sign Language dialect, perform deep learning based Sign Language Recognition modeling, and report the benchmark performance. The Bornil platform, BornilDB v1.0 Dataset, and the codebases are available on https://bornil.bengali.ai.

## 1 Introduction

Despite the presence of more than 300 sign language dialects in use globally, the establishment of a universal or lingua-franca of sign language remains elusive. According to the World Health Organization (WHO), there is a substantial population of 430 million individuals with auditory/hearing impairments, including 34 million children (WHO, 2021). It is noteworthy that 72 million deaf individuals worldwide rely on sign language as their primary means of communication (UN2, 2019), while for many others, including family members and caregivers, it serves as a secondary language.

Since sign language is the medium of communication for two types of disabilities, prevalence of its use helps better integration in society for a large number of people. Of the different sign language dialects practiced in the world, very few have been properly documented, and only a handful are moving toward complete digitization (Strobel et al., 2023).

While there has been scientific documentation and advancement in the field of Automatic Sign Language Recognition (Koller, 2020), Generation (Bragg et al., 2019), and associated technologies (Duarte et al., 2022; Ventura et al., 2020) for prominent sign language dialects such as American Sign Language (ASL) and British Sign Language (BSL), this progress is primarily attributed to the availability of extensive datasets and corpora. In contrast, many other sign language dialects lack similar attention and resources (Polat and Saraçlar, 2020), rendering them as zero or low-resource sign language dialects in terms of automation and digitization.

Sign languages encompass a rich array of ele-

---

* Denotes equal contributions

*Demo for the Bornil platform is available at: Youtube Link

| Dataset Source | Year | Language | Language Level | Duration/Samples | Signers | Multiview | Transcription | Gloss | Pose |
|---|---|---|---|---|---|---|---|---|---|
| Duarte et al. (2021) | 2021 | ASL | Continuous | 80 Hours | 11 | ✓ | ✓ | ✓ | ✓ |
| Mukushev et al. (2022) | 2022 | KRSL | Continuous | N/A | 50 | X | ✓ | ✓ | X |
| Agris and Kraiss (2010) | 2010 | DGS | Continuous | 55 Hours | 25 | X | ✓ | ✓ | X |
| Camgoz et al. (2018) | 2018 | DGS | Continuous | 11 Hours | 9 | X | ✓ | ✓ | X |
| Huang et al. (2018) | 2015 | CSL | Continuous | 25,000 Samples | 50 | X | ✓ | ✓ | X |
| Nandy et al. (2010b,a) | 2009 | ISL | Isolated | 605 Samples | N/A | ✓ | X | ✓ | X |
| Kapuscinski et al. (2015) | 2015 | PSL | Isolated | 1,680 Samples | 1 | ✓ | ✓ | ✓ | X |
| Aran et al. (2007) | 2007 | TSL | Isolated | 440 Samples | 11 | X | ✓ | ✓ | X |
| Ronchetti et al. (2016) | 2016 | LSA | Isolated | 3,200 Samples | 10 | X | ✓ | ✓ | X |
| **BornilDB v1.0** | **2023** | **BdSL** | **Continuous** | **21,154 samples** | **3** | ✓ | ✓ | ✓ | ✓ |

**multiview:** recorded from multiple angles; **transcription:** text representation of the content of the video, pose: body-face-hands keypoints
**gloss:** transcription of signs using spoken language words indicating what individual parts of each sign mean ments, comprising gestures (Liddell, 2003) characterized by handshape, orientation, movement, and locations, as well as expressions, occasionally accompanied by mouthing (Nadolske and Rosenstock, 2007; Albanie et al., 2020). Consequently, the automatic recognition of sign language emerges as a dynamic and formidable research domain (Rastgoo et al., 2021; Camgoz et al., 2020), owing to the multifaceted visual and linguistic hurdles it entails. These challenges pose significant obstacles in the realm of robust feature extraction and modeling (Bragg et al., 2019), further accentuating the complexity of the field.

Table 1: Summary of publicly available international continuous sign language datasets

Even though Sign languages are capable of conveying the same meanings as spoken languages do, sign language dialects have grammatical structures independent from the associated spoken language (Sandler, 2010). Also, there is linguistic diversity within every sign language dialect, just like spoken languages (Braithwaite, 2019). Although recent deep learning advances aim towards robust and explainable feature extraction (Bai et al., 2021); to ensure proper optimization, we need to collect a large volume of data that is representative of the real-world distribution to ensure good enough performance in-the-wild (Stadelmann et al., 2018). These are, albeit general criticisms of deep learning, an even more prominent issue impeding the progress of Automated Sign Language Recognition (ASLR). Within this realm, the acquisition of a substantial and diverse dataset poses considerable challenges, stemming from the absence of established frameworks for data collection, exorbitant logistical costs, and the scarcity of diversified sources for unannotated data. Consequently, the growth of sign language research has not paralleled the advancements observed in similar fields; such as speech recognition (Li et al., 2022; Roger et al., 2022). In an effort to overcome these obstacles and foster accelerated progress, we have developed *Bornil*, a pioneering platform for crowdsourcing sign language video data. This platform facilitates the collection and curation of both scripted and spontaneous sign language data across various sign language dialects, thereby providing a robust foundation for further research endeavors.

Through Bornil, we enable the creation of both sentence-level and gloss-level crowdsourcing of video datasets for Sign Language encompassing over 300 low-resource dialects. The intended application of the dataset is to facilitate Automated Sign Language Recognition (ASLR) of the aforementioned dialects, with the ultimate goal of promoting efficient communication for individuals using sign language. To democratize the data collection process and expedite the development of ASLR, we offer this open-source platform to address multiple data collection process factors to ensure data collection for all sign language dialects.

## 2 Challenges with Existing Datasets & Tools

### 2.1 Existing Datasets

A number of continuous sign language datasets have been collected over the years. We show some of these different sign language datasets in Table 1 However, many of these datasets come from expensive projects, and some require studio environments that are not suitable for low-resource sign languages (Jiang, 2022; Uthus et al., 2023). Due to a scarcity of annotated diversified datasets, many sign language dialects remain zero-resource. Also, the lack of a proper data collection platform makes it very hard to collect data through crowdsourcing. Due to this, many sign language data collection initiatives could not scale; leading to a scarcity of large-scale datasets.

For Bangladeshi sign language recognition, datasets that exist are only image datasets (Rafi et al., 2019; Ahmed and Akhand, 2016; Yasir et al., 2015; Islam et al., 2018; Uddin and Chowdhury, 2016; Podder et al., 2020; Hasan et al., 2021). These datasets focus on only the representation of various Bangla alphabets using hand gestures. Tazalli et al. (2022), however, provides a video dataset. But, each of these videos is of a single word and not a complete sentence. Also, similar to the image datasets, this one also focuses on hand movement only and ignores the other important non-manual markers of sign language e.g. movements of the head (nod/shake/tilt), mouth (mouthing), eyebrows, cheeks, facial grammar (or facial expressions) and eye gaze. Promising recent works in sign language processing have shown that modern computer vision and machine learning architectures can help break down these barriers for sign language users.

## 2.2 Existing Platforms & Tools

While ideas and frameworks exist for sign language data crowdsourcing platforms (Farooq et al., 2021), there are hardly publicly available usable ones. Annotation tools such as Elan (Sloetjes and Wittenburg, 2008), SignStream (Neidle et al., 2001), iLex (Hanke, 2002) etc. are very helpful in creating annotations. However, when it comes to crowdsourcing, having a single platform with all the necessary tools makes it easier to maintain and collect data rather than using multiple tools, be it in-house or of-the-shelf, for different purposes.

## 3 Platform Description

Our platform, *Bornil*, is targeted towards video-based continuous sign language dataset collection and annotation. It provides features like data collection, validation, and annotation for multiple languages through crowdsourcing (inspired by Mozilla Common Voice (Ardila et al., 2019)); and the recordings do not have restrictions like greenscreens or noise removal. Recordings are saved with all noises that are present during recording to ensure that they represent real-life scenarios. The platform is usable on an array of devices, thus removing the need for a specific setup, which is inconvenient for crowdsourcing. Dataset created through the platform includes body-face-hands keypoints generated manually using openpose (Cao et al., 2019; Simon et al., 2017; Cao et al., 2017; Wei et al., 2016) & mediapipe (Lugaresi et al., 2019).

Our platform provides four main features; **Video Recording, Validation, Annotation** and **Multilingual support.** We provide support for both sentence-level and gloss-level annotation. Validation is further broken down into Video Validation (which includes metadata validation) and Annotation Validation. We also collect essential metadata like gender, age, locality, etc., for diversified data collection and stratified analyses.

One major focus that we had while developing the platform is its general ease of use. Bornil has an intuitive interface, which enables data collection from a wide array of sources, including the Deaf community, hard-of-hearing individuals, children of dead parents, and siblings of deaf adults. To offer real-world representative data, voluntary contributors may assist by adjusting camera resolution, background noise, and other metadata.

A detailed description of the workflow of our system is given in the following subsections while a summary workflow is shown in Figure 2. Figure 3, 4, 5 show snippets of our recording, validation, and annotation features.

### 3.1 Upload Text/Topic

Texts or topics must be uploaded to the server by an admin for which users create sign-language video recordings. This is done by uploading a CSV file. The file consists of 3 columns; content, content type (text or topic), and language. During the recording, users are assigned texts at random.

### 3.2 Video Recording

In the video recording section, the user sees a built-in video recorder. (S)he is assigned a text (one or more sentences or a topic) based on his/her selected language for which (s)he will record a video in sign language. If a text is assigned, the user will record herself signing that text. If a topic is assigned, the user will record herself signing a preset number of sentences on that topic.

After recording the video, the platform automatically switches to a video player from where a user can preview the recording. The user can then set the start time and end time to remove the unnecessary parts (such as moving the hand forward to start or end the recording). If a mistake is made during recording, the user has the choice of keeping the video or discarding it. If the video is discarded, (s)he is prompted to record the video again. Otherwise, a form is provided where (s)he must fill up video details such as lighting information, camera view, resolution, etc.

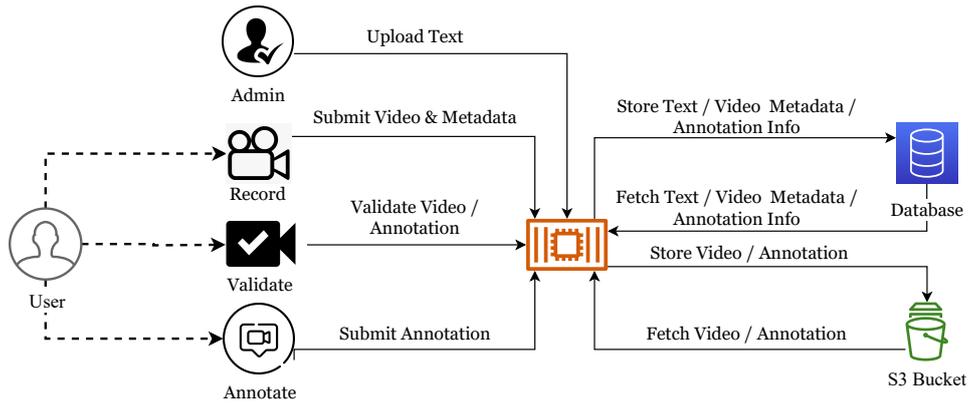

Figure 2: High-Level Workflow Diagram of the platform.

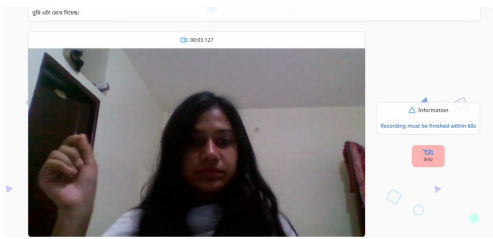

Figure 3: Video Recording

After submitting all the required fields, Bornil uploads the video to the server and stores the information against the user's account. User is given the option to annotate the video– which is optional.

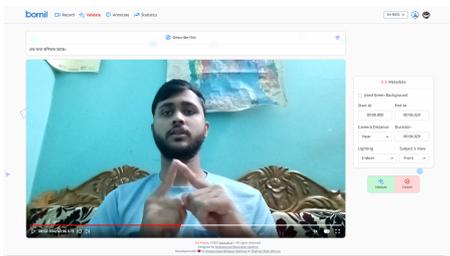

Figure 4: Video Validation

### 3.3 Video Validation

In the validation section, user is provided with a video, its corresponding text, and other information provided during the video recording section. (S)he then plays the video and provides feedback on whether the video is correct or not in accordance with the text. (S)he can also edit the start time, end time, view, lighting, and other information if (s)he finds it wrong.

### 3.4 Video Annotation

This section gives users an unannotated video and its corresponding text/topic. User is given an interface with text boxes and timestamps which (s)he will use for the annotation. The platform provides an interactive timeline with a transcription edit feature, allowing the user to freely change the length, start time, and end time of the text boxes and also seek the video to a specific timestamp.

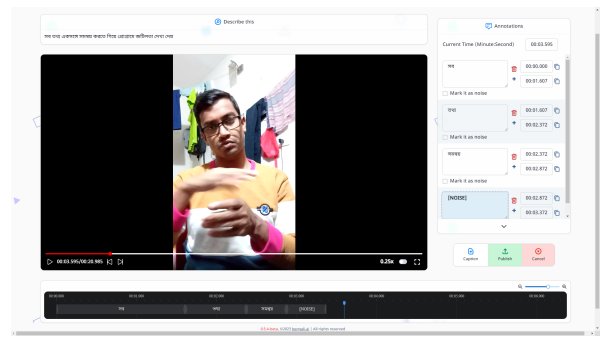

Figure 5: Video Annotation

Annotation will be somewhat different based on whether the video was recorded for a text or a topic. If the video is for text, the user will be shown the text along with the video. However, if the video was on a topic, the user must first type/write down the script for which the video was recorded. For sentence-level annotation, user sets the timestamp for each sentence in the video. For gloss-level annotation, the user sets the timestamp for each word in the video.

### 3.5 Annotation Validation

After annotation has been added for a video, it will appear in this section. Here, annotations will be shown as subtitles based on their timestamps. During this validation, User is given the option to fix any annotation-related errors that have been found. For updating the annotation, user is given a similar interface with text boxes and timestamps used during the video annotation process.

### 3.6 Accessibility Consideration

**Multilingual Support**

Bornil is equipped with the necessary technical capabilities to facilitate data collection for multiple distinct sign languages, as well as various interface languages. However, at present, we are in the process of launching our website with a focus on providing support for en-ASL (English to American Sign Language) and bn-BdSL (Bangla to Bangladeshi Sign Language) translations. We plan to gradually introduce support for additional languages based on user demand in the future.

**Tutorial Support**

For ease of use, we have included a video tutorial on the homepage of our website, covering all major features. If users encounter difficulties while accessing these features, they can refer back to the tutorial. The website has been designed to be intuitive and simple, ensuring accessibility for all.

**System Requirements**

Our platform possesses a lightweight design and is compatible with devices costing as low as $120. Extensive testing has been conducted on this device, affirming that all features function seamlessly with minimal complications.

## 4 Tech Stack Description

### 4.1 Frontend

We prioritized open-source frameworks with stability for development. Our proposed system is modular, easily deployable, and customizable. For the front end, NextJS with Typescript support is adopted. NextJS enables server-side rendering while loading JS and style files only for the requested pages. This rendering process significantly reduces the data needed to render any page. Moreover, it supports cached static HTML pages to be served to the user, significantly decreasing the load time for most users.

### 4.2 Backend

We used Go with its built-in HTTP server for the backend. Go is a statically typed, compiled programming language. Since Bornil is a crowdsourcing platform, using concurrency features such as Goroutine and channels to process many user requests asynchronously is necessary. As a video data collection platform, Go's garbage collection and performance efficiency effectively handles the large amount of data we receive. PostgreSQL is chosen as Bornil's database management system because of its flexibility. Parallelization of data queries while maintaining data integrity makes it an excellent candidate for this project. In addition, we have used Amazon S3 (Amazon Simple Storage Service) as our video storage service. Since our system requires an affordable, scalable storage solution with high data availability, we have opted to go with Amazon S3, which offers 99.9% data availability and cheaper data plans compared to its contemporary competitors.

## 5 System Efficacy

| Sentence Length | Samples | BLEU1 | BLEU2 | BLEU3 | BLEU4 |
|---|---|---|---|---|---|
| 5 | 1146 | 12.01 | 5.35 | 2.58 | 1.3 |
| 7 | 1728 | 10.19 | 4.08 | 1.75 | 0.79 |
| 10 | 2110 | 8.66 | 3.36 | 1.36 | 0.59 |

Table 2: Performance of the test set on data collected from Bornil

To demonstrate the efficacy and the effectiveness of the proposed data crowd-sourcing platform, we present the platform to a test group of 3 sign language literate people with congenital hearing loss. All the metadata have been stored along with each data recording using the platform.

The control group contributed towards the development of the "BornilDB v1.0" which consists of 73 hours of sign language video data, including 21,104 scripted sample and 50 spontaneous recordings where a word-based topic stimulus was presented on the screen. BornilDB v1.0 consists of 138,586 total and 25,572 unique Bengali words. The average Bengali word count per recording is 7.704, with an average video duration of 13.374 seconds. In less than three months, the control

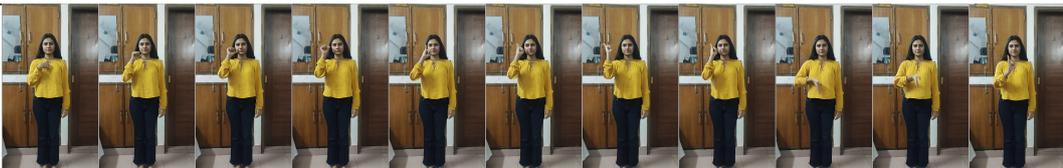

Figure 6: Preview of The Dataset Created Using the Platform. আমি (Pron:1st), আগামীকাল (Adv: tomorrow) বেড়াতে (trip: verb root বেড়ানো + suffix "-তে") যাবো (Go: verb root "যাওয়া" + "বো" First Person Singular; Future)

group led to the development of the 2nd largest dataset of its kind, compared to the other deep learning-based sentence-level video datasets in Table 1. We also finetuned a transformer model pretrained on How2Sign dataset (Tarrés et al., 2023) on our collected data and present our baseline result on table 2. We believe that this establishes the potential of this crowdsourcing platform in the creation of deep learning datasets for the different sign language dialects.

Figure 6 shows an example of the dataset with mediapipe & openpose keypoints.

## 6 Conclusion

Through this work, we present the first comprehensive deep-learning data crowdsourcing platform for ASLR. We utilize an industry-grade development practice in our technological stack. Our platform provides an intuitive and inclusive interface to collect data from a wide array of data contributors. We provide interactive data annotation and validation facilities for smooth adaptation. Moreover, we have shown our platform's efficacy by collecting its second-largest crowdsourced sentence-level video sign language dataset. We will continue the data crowdsourcing efforts to provide deep learning practitioners with reliable and interpretable datasets for different sign language dialects.

## 7 Limitations

Efforts are currently underway to achieve optimal compatibility with the Safari browser and Apple WebKit engine. We have listed the devices and browsers that are currently compatible with our platform in the appendix. This list will be expanded as we introduce more compatibility. Poor network connection may occasionally precipitate long video upload durations, thereby culminating in server timeouts, irrespective of the platforms involved.

## Ethics and Broader Impact Statement

**Data** We take proper consent from the data contributors through the platform regarding the distribution and use of the video recordings and other metadata for research purposes. We release the datasets regularly under CC by SA 4.0 license.

**Impact**

Bornil crowdsourcing platform can be used to crowdsource and curate high-quality ASLR datasets for all the major sign language dialects. Due to the design of the platform, a large group can collaborate in creation of resources. Through our data crowdsourcing pilot initiative, we have shown that this platform is already effective for the creation of sign language recognition datasets and successful modeling 3 can be done on the data. Hence, we are providing a complete framework for digitizing any sign language dialect.

The system is not resource hungry and can be accessed through any personal computer or even cellphones. This, furthermore, helps in scaling the data crowdsourcing process.

It is evident that this platform has the potential to bridge the communication gap between the deaf and the general public. We make the system available free of cost for research, and publish our codes and data regularly.

## A  Appendix

Our platform has been extensively tested on the following devices & browsers. We are working towards providing support for more devices & browsers.

Tested browsers:

- Google Chrome: Desktop version 109.0.5414.119 (Official Build) (arm64)
- Google Chrome: Android version 110.0.5481.65
- Brave: Desktop version 1.48.164 Chromium: 110.0.5481.100 (Official Build) (arm64)
- Firefox: Desktop version 109.0.1 (arm64)

Tested devices:

- Macbook Pro 14 inch (2021 base model)
- Redmi Note 7 Pro
- Redmi Note 5A Prime

Tested operating systems (OS):

- MacOS Ventura (arm64)
- Windows 10 (amd64)
- Ubuntu 22.04.1 LTS (Kernel version 5.15.0-58-generic)
- Android 7.1.2
- Android 10 (Android Q)

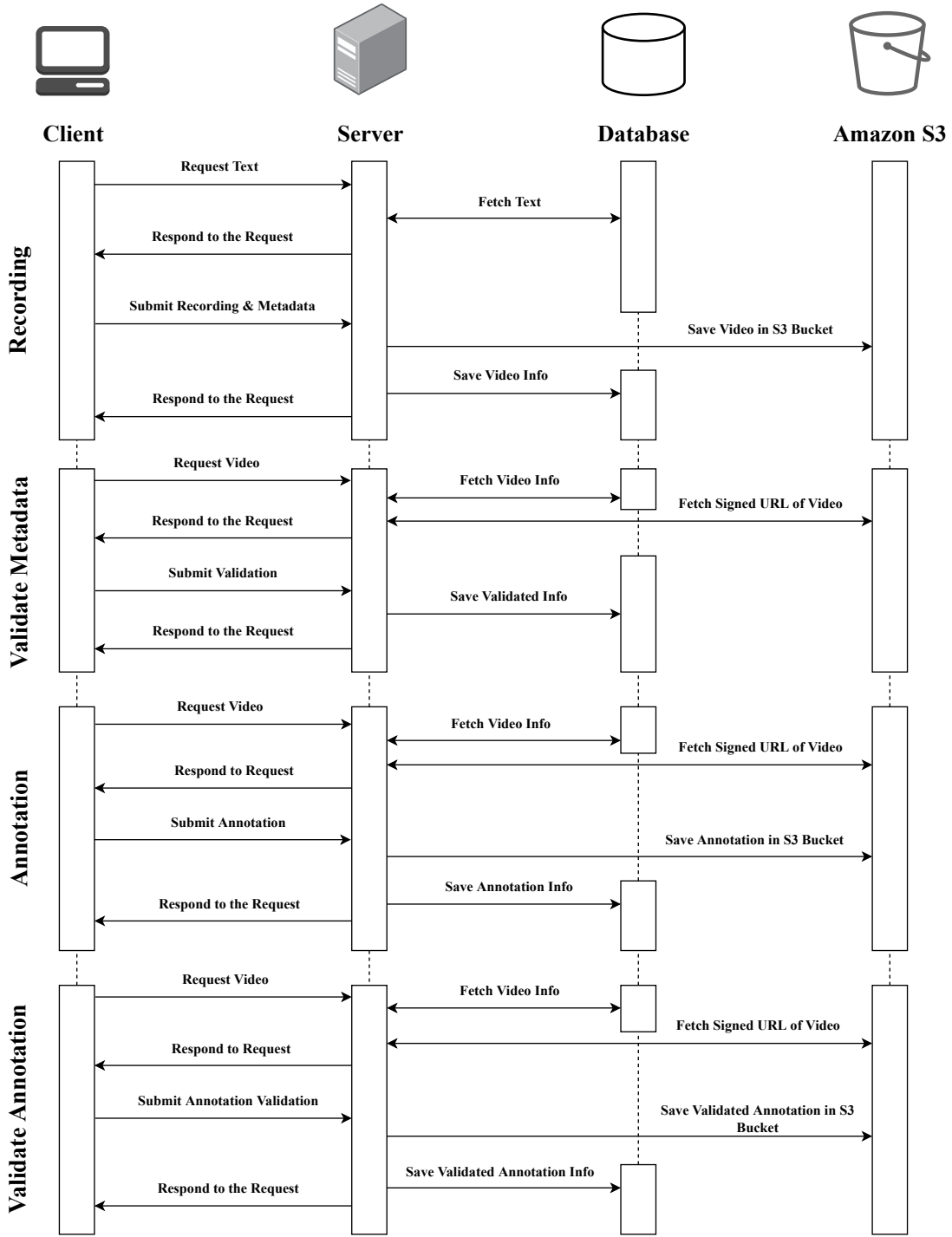

Figure 7: Detailed workflow Diagram of the platform

| Reference | Prediction |
| --- | --- |
| সব কিছু ঠিক আছে তো? | সব ঠিক আছে? |
| কিন্তু এমন টা তো হতে দেখা যাচ্ছে না। | কিন্তু এমন কাওকে দেখা যায় না। |
| না, আমার মনে হয় না। | না, আমার মনে হয় নেই। |
| আচ্ছা জলদি শুরু করো। | না, চলো শুরু করি। |
| তাই অপেক্ষা করিনি। | আর অপেক্ষা করুন। |

Table 3: Model Prediction for smaller sentences. This shows the efficacy of the system.